\documentclass[a4paper,USenglish,cleveref, autoref, thm-restate]{oasics-v2021}

\nolinenumbers

\graphicspath{{./figures/}}

\title{Bias in Knowledge Graphs -- an Empirical Study with Movie Recommendation and Different Language Editions of DBpedia}

\titlerunning{Bias in Knowledge Graphs}

\author{Michael Matthias Voit}{University of Mannheim, Germany}{michael.matthias.voit@gmail.com}{}{}
\author{Heiko Paulheim}{University of Mannheim, Germany\and \url{http://www.heikopaulheim.com}}{heiko@informatik.uni-mannheim.de}{https://orcid.org/0000-0003-4386-8195}{}

\authorrunning{Voit, M. M. and Paulheim, H.} 

\Copyright{Michael Matthias Voit and Heiko Paulheim} 



\EventEditors{John Q. Open and Joan R. Access}
\EventNoEds{2}
\EventLongTitle{Language, Data and Knowledge (LDK 2021)}
\EventShortTitle{LDK 2021}
\EventAcronym{LDK}
\EventYear{2021}
\EventDate{June 14-16, 2021}
\EventLocation{Zaragoza, Spain}
\EventLogo{}
\SeriesVolume{42}
\ArticleNo{23}

\begin{document}

\maketitle

\begin{abstract}
Public knowledge graphs such as DBpedia and Wikidata have been recognized as interesting sources of background knowledge to build content-based recommender systems. They can be used to add information about the items to be recommended and links between those. While quite a few approaches for exploiting knowledge graphs have been proposed, most of them aim at optimizing the recommendation strategy while using a \emph{fixed} knowledge graph. In this paper, we take a different approach, i.e., we fix the recommendation strategy and observe changes when using different underlying knowledge graphs. Particularly, we use different \emph{language editions} of DBpedia. We show that the usage of different knowledge graphs does not only lead to differently biased recommender systems, but also to recommender systems that differ in performance for particular fields of recommendations.
\end{abstract}

\section{Introduction}
Large-scale knowledge graphs, such as DBpedia \cite{lehmann2015dbpedia} and Wikidata \cite{vrandevcic2014wikidata}, are recognized as a valuable ingredient for intelligent applications \cite{heist2020knowledge}. In such applications, they can provide background information on the entities processed, which often leads to performance improvements in downstream processing steps \cite{ristoski2016semantic}.

In the past, different works have been proposed on building recommender systems based on knowledge graphs, most prominently, DBpedia. The first of those approaches has probably been \emph{dbrec}, dating back to 2010 \cite{passant2010dbrec}. Since then, a number of approaches have been proposed, challenges around the topic have been conducted \cite{di2014linked}, and recent approaches have been utilizing the omnipresent knowledge graph embeddings for computing recommendations \cite{musto2019embedding,rosati2016rdf}.

The vast majority of those works always utilizes a fixed knowledge graph (DBpedia in most cases) and then optimizes the recommendation algorithm to provide the best empirical results on a test dataset. This means that by fixing the knowledge graph upfront, the influence of the chosen graph, its coverage, data quality, and possible biases, are not examined.

In this paper, we postulate that the choice of a particular knowledge graph has an influence on the behavior of the overall system, and may lead to a certain bias. To analyze this bias, we train a  recommendation system with a fixed setup and parameter settings based on the embedding method RDF2vec~\cite{ristoski2019rdf2vec}, using different versions of DBpedia, which have been extracted from Wikipedia language editions.

Assuming that the coverage, quality, and level of detail of recommended items (in our test scenario: movies) varies from language edition to language edition, we expect a certain bias to shine up when using different knowledge graphs. This is confirmed by our experiments, however, the bias is not as obvious as we expected. While the straight forward assumption is that, e.g., a recommender system based on the German DBpedia edition would develop a stronger bias towards recommending German films, the effects are more subtle than that, exposing different significant biases with respect to production country and genre.

The rest of this paper is structured as follows. Section~\ref{sec:related_work} discusses related works. In section~\ref{sec:experiments}, we lay out our experiment set up, followed by an analysis of findings in section~\ref{sec:findings}. We conclude with a summary and an outlook on future work.

\section{Related Work}
\label{sec:related_work}

The two most well-known families of recommender systems are collaborative filtering and content-based recommender systems. While the former analyze the behavior of users and recommends items that are consumed by users with a similar behavior as the one at hand, the latter exploit similarities between the items per se. For that category of approaches, a model of the recommended items is required, which can be unstructured (e.g., a textual description) or structured (e.g., a set of attributes). \cite{ricci2015recommender}

For structured representations, public knowledge graphs like DBpedia or Wikidata have been recognized as a valuable source of information, since they already contain a large amount of information on various items in a structured form \cite{heist2020knowledge}. Most classic approaches use  DBpedia and/or knowledge graphs tailored to the domain at hand, and base their decision on a similarity function based on a set of hand-picked attributes (e.g., genre and artist for music; genre, director, and actor for movies).

The first generation of recommender systems based on Knowledge Graphs, such as dbrec, were based on hand-picking attributes and relations. Later approaches also exploited automatic approaches for selecting attributes, either by adapting measures such as TF-IDF to graph data \cite{di2012exploiting}, or by using machine learning methods such as Random Forests, which can be used on larger feature sets and automatically identify the relevant ones \cite{ristoski2014hybrid}.  

\begin{figure}[t]
    \centering
    \includegraphics[width=0.75\textwidth]{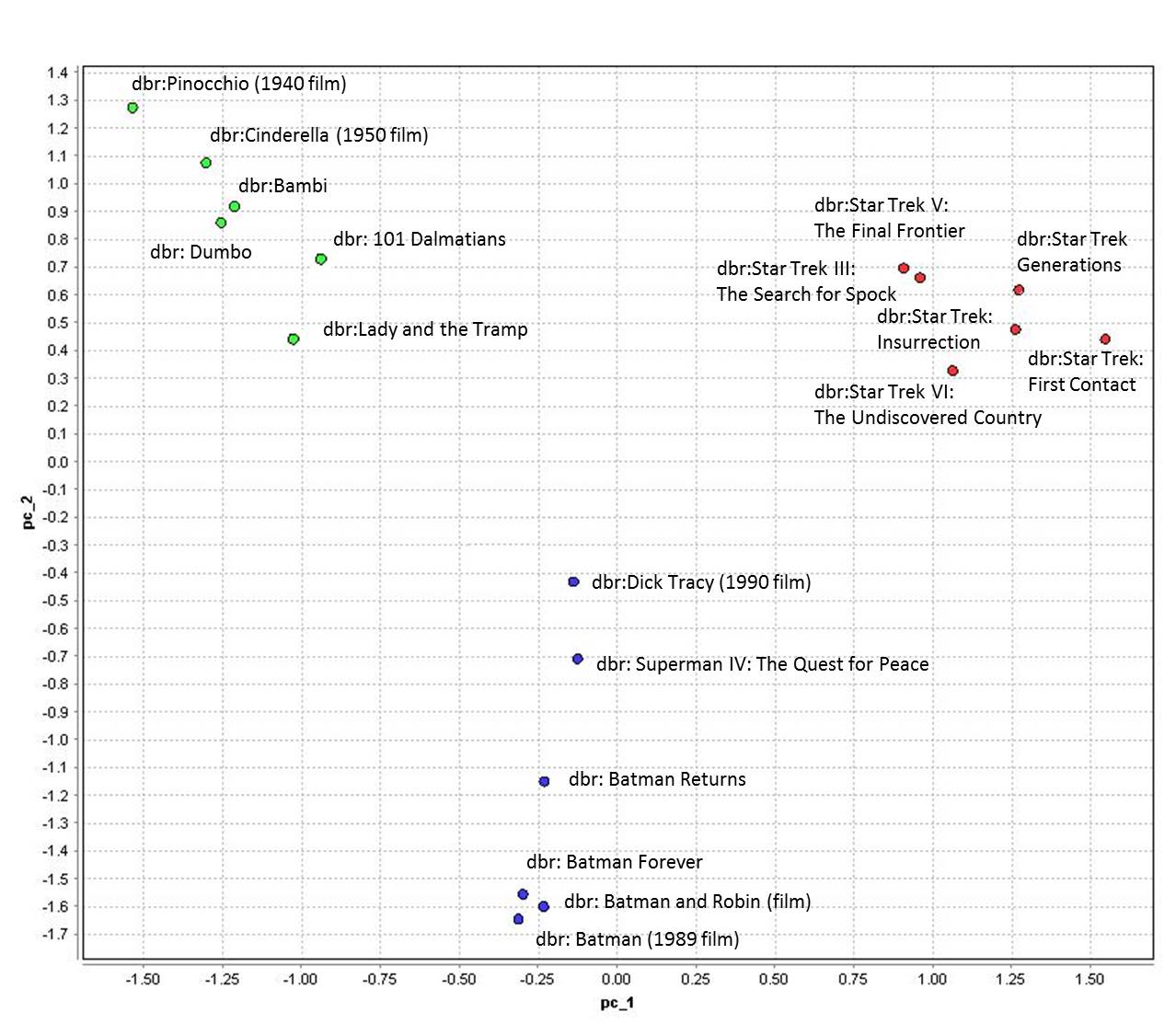}
    \caption{2-dimensional PCA projection of embedding vectors for a set of movies in DBpedia \cite{rosati2016rdf}}
    \label{fig:example_pca}
\end{figure}

The most recent generation of such recommender system utilizes \emph{knowledge graph embeddings} \cite{guo2020survey}. Such embedding methods project resources in a knowledge graph into a lower-dimensional, numerical vector space. Since many of those projection methods lead to vector spaces in which similar resources are close to each other, distance in the embedding space can be exploited for recommendation, as depicted in figure~\ref{fig:example_pca}. Such approaches, among others, have been analyzed for RDF2vec \cite{rosati2016rdf}, metapath2vec \cite{yang2018knowledge}, 
TransE \cite{dadoun2019location,huang2019mpd,huang2018improving,wang2018dkn,wang2019herb}, 
TransR \cite{lin2018heterogeneous,tang2019akupm,wang2018dkn,wang2017safe,zhang2016collaborative}, 
TransH \cite{cao2019unifying,wang2018dkn}, 
TransD \cite{he2019hi2rec,wang2018dkn}, 
ComplEx \cite{huang2019mpd}, LINE \cite{wang2017safe}, Laplacian Eigenmaps and node2vec \cite{musto2019embedding,palumbo2017entity2rec}, and embedding methods specifically tailored to the recommendation task, like \emph{RippleNet} \cite{wang2018ripplenet}, \emph{CFKG} \cite{zhang2018learning},  \emph{Hierarchical Collaborative Embedding} \cite{zhou2018knowledge}, \emph{MKR} \cite{wang2019multi}, and \emph{UPM} \cite{zhu2019neural}. More recently, graph neural networks have also gained a bit of traction \cite{wang2019knowledge2,wang2019knowledge}.

While we are aware that this set of examples for the usage of knowledge graphs for content-based recommender systems is far from complete, a common trend can be observed in almost all publications about such systems: they always fix the knowledge graph to be used upfront, and different variants are typically studied for the algorithms used to compute similarities, but not for the graph as such. In the rare cases where results obtained with multiple knowledge graphs are examined (e.g., \cite{rosati2016rdf}, which contrasts results based on DBpedia and Wikidata), they are only compared based on the scoring function (e.g., F1 score), but other influences on the behavior of the recommender system are not analyzed. Hence, the influence of the choice for a particular knowledge graph is still underexplored.

In this paper, we conduct a study to shed some light on that aspect. To that end, we use different versions of DBpedia. DBpedia is extracted from Wikipedia infoboxes by the use of mappings to a central ontology. There are versions of DBpedia for different Wikipedia languages, called \emph{DBpedia language editions} \cite{lehmann2015dbpedia}.

It is known that language editions of Wikipedia differ in coverage and level of detail. Their size ranges from a few hundred to a few million articles.\footnote{\url{https://en.wikipedia.org/wiki/List_of_Wikipedias}} These differences have been analyzed with respect to various aspects, e.g., topical coverage \cite{bao2012omnipedia,eom2015interactions,hecht2009measuring,miz2020trending}, article quality \cite{lewoniewski2016quality} and neutrality \cite{callahan2011cultural,massa2012manypedia,wessler2017wikiganda,zhou2016likes},  bias related to geography \cite{beytia2020positioning} or gender \cite{wagner2015s}, and user behavior \cite{hara2010cross,lemmerich2019world}.

The difference in the quality of infobox data in different Wikipedia language editions has also been studied \cite{wkecel2015modelling}. This is particularly interesting for our scenario, since the DBpedia knowledge graph draws its information from those infoboxes. Hence, in the light of those studies, we expect significant differences in knowledge graphs extracted from different Wikipedia languages, and we want to explore in how far they lead to difference in downstream applications such as recommender systems.

\section{Experimental Setup}
\label{sec:experiments}
In our experiments, we use the MovieLens 1M dataset \cite{harper2015movielens}, which contains a 1M 1-5 star ratings by 6,040 users for 3,952 movies. Moreover, we use DBpedia version 2016-10.\footnote{\url{https://wiki.dbpedia.org/develop/datasets/dbpedia-version-2016-10}}

In earlier works, links from MovieLens 1M to DBpedia have been provided \cite{noia2016sprank}. Since DBpedia has been evolving since the original linking was performed, we removed all links that refer either to entities which do not exist anymore (i.e., the URI has changed in later releases of DBpedia) or to entities derived from disambiguation pages. Our resulting dataset consists of 3,123 movies linked to the English DBpedia.

\subsection{Datasets}

\begin{table}[t]
    \begin{tabular}{|p{2.5cm}||c|c|c|c|c|c|c|c|c|c|}
        \hline
         & \textbf{it} & \textbf{pl} & \textbf{es} & \textbf{pt} & \textbf{fr} & \textbf{de} & \textbf{ru} & \textbf{nl} & \textbf{ja} \\
        \hline
        \textbf{\# Movies} & \textbf{24k} & 13k & 12k & 12k & \textbf{16k} & \textbf{19k} & \textbf{15k} & 10k & 10k \\
        \hline
        \textbf{Intersection with MovieLens 1M mapped to DBpedia-en} & \textbf{2,610} & 2.106 & 2,019 & 2,092 & \textbf{2,658} & \textbf{2,426} & \textbf{2,255} & 1,793 & 1,888 \\
        \hline
    \end{tabular}
    \caption{Statistic of common movies in different language version with the movies mapped to the English DBpedia}
    \label{tab:language_movies}
\end{table}

To investigate the influence of the usage of different versions of DBpedia on a recommender system, we utilize different language versions of DBpedia. In a preliminary study, we looked at the ten largest language editions of DBpedia\footnote{\url{https://wiki.dbpedia.org/services-resources/datasets/dataset-statistics}}, and analyzed the overlap with the 3,123 movies in our dataset linked to the English DBpedia. To that end, we utilze the links between DBpedia versions, which are extracted from the inter-language links in Wikipedia. The results are depicted in table~\ref{tab:language_movies}.

To ensure a reasonable coverage, we decided to use the five dataset which have the most information about movies and the highest overlap with the English dataset. Hence, we decided to base our analysis on English, Italian, French, German, and Russian. The subset of the original 3,123 movies which have a corresponding entity in all five datasets comprises 1,948 movies.

We apply two additional filtering steps, as suggested by \cite{cremonesi2010performance}, \cite{noia2016sprank}, and \cite{ristoski2019rdf2vec}. To avoid a popularity bias, the top-rated 1\% of all movies are removed. In the second step, users with less than 50 ratings are removed, and so are movies without any ratings. After this step, we obtain a dataset with 1,918 movies, 675,960 ratings, and 3,642 users. This set is used as the basis for all our experiments.

\subsection{Recommender Algorithm}
As discussed above, in our experiments, we aim at keeping the recommender algorithm fixed, while varying the underlying knowledge graph. We intentionally use a simple algorithm for the recommendations, as our goal is not to maximize the performance of the recommendation as such, but to examine the influence of the underlying knowledge graph.

Following the setup in \cite{ristoski2019rdf2vec}, we use RDF2vec to compute vector space embeddings of the different DBpedia graphs. RDF2vec extracts random walks from a knowledge graph, which are represented as ``sentences'' of entities and predicates in the knowledge graph. On that set of sentences, the word2vec algorithm \cite{mikolov2013distributed} is run, which then computes an embedding vector for each entity (and predicate).

We computed RDF2vec embeddings for the five DBpedia language editions identified above, using the best performing parameter setting identified in \cite{ristoski2019rdf2vec}, i.e., extracting 500 walks per entity with a depth of 4, a dimensionality of 200 for the word2vec model, using the Skip-Gram variant.\footnote{All code and data is available online at \url{https://github.com/voitijaner/Movie-RSs-Master-Thesis-Submission-Voit}}

The similarity of two movies is then computed as the cosine similarity of the corresponding vectors in that vector space. To that end, we compute a score \(y_{uj}\) for an unrated movie \(j\) and user \(u\) is calculated with the following formula:

\begin{equation}
y_{uj}  = \frac{\sum \nolimits_{i \in I_{u} } cos(i,j) * r_{ui}}{\sum \nolimits_{i \in I_{u} } cos(i,j) }
\label{form:score_list}
\end{equation}

Here, \(I_{u}\) are the previous observations from user \(u\) and \(r_{ui}\) denotes the rating of item \(i\) of the user \(u\) in the training set.
Then, the N movies with the highest scores are returned for each user. For this procedure, we used the item similarity recommender of the GraphLab Create python framework\footnote{\url{https://turi.com/products/create/docs/generated/graphlab.recommender.item_similarity_recommender.ItemSimilarityRecommender.html}}.

\subsection{Metrics}
To evaluate the quality of recommendations, we use the standard measures of recall, precision, and F1 score. Here, recall measures the fraction of items that a user ultimately rated positively which were recommended to him or her, and precision measures the fraction of recommended items which were ultimately rated positively. F1 is the harmonic mean between the two.

Besides the quality of recommendations, we are interested in differences among recommendations created based on different knowledge graphs. To that end, we look at different \emph{categorical features}, like language or genre. For a categorical feature $C$ (such as \emph{language}), we can compute the probability of recommendations with a certain feature value $c$ (such as \emph{German}), i.e.,
\begin{equation}
    p(c) = \frac{\left|R_c\right|}{\left|R\right|}
\end{equation}
where $\left|R_c\right|$ is the total number of items that were recommended by a certain approach which have the categorical feature $c$, and $\left|R\right|$ is the total number of recommendations computed. These probabilities can then be compared for recommender systems based on different knowledge graphs.

In order to distinguish random variations from effects actually induced by the use of different knowledge graphs, we additionally conduct a chi-squared test:
\begin{equation}
\chi^2_{KG}=  \sum \nolimits_{c \in C} \frac{(|R_{c}| - (c_e * |R|))^2}{(c_e * |R|)}
\end{equation}
where $c_e$ is the expected value of recommendations with a categorical feature value $c$. We sum up the $\chi^2_{KG}$ values for all KGs, and compare them against the $\chi^2$ distribution with $(\left|KGs\right|-1)\cdot(\left|C\right|-1)$ degrees of freedom, and an $\alpha$ value of 0.05 to test for significance. All the results presented below are not a random result according to that definition.

\section{Findings and Observations}
\label{sec:findings}
In total, we compare five recommender systems, based on the five different knowledge graphs. We analyze both the overall performance, as well as biases w.r.t. production countries and genres.
\subsection{Overall Performance}
\begin{table}[t]
    \centering
    \begin{tabular}{|l|r|c|c|}
        \hline
        \textbf{KG} & \textbf{Precision} & \textbf{Recall} & \textbf{F1 score} \\
        \hline
        de & 0.057 & 0,0404 & 0.047 \\
        fr & 0.054 & 0.038 & 0.044 \\
        en & 0.053 & 0.038 & 0.044 \\
        it & 0.048 & 0.036 & 0.042 \\
        ru & 0.042 & 0.028 & 0.034 \\
        \hline
    \end{tabular}
    \caption{Performance of the recommender systems per KG}
    \label{tab:overallPerformance}
\end{table}
In a first analysis, we look at the overall performance difference between the recommenders based on the five knowledge graphs. We can see that the one based on the German DBpedia works best, which is most likely due to a higher linkage degree for movies.

Most strikingly, the English DBpedia -- which is used as the basis for the majority of works which claim to use ``DBpedia'' as a source of background knowledge -- performs worse than its German and French counterpart. This shows that this choice, which is often done by simple heuristics such as popularity and availability, might not be an optimal one.

\subsection{Bias for Production Countries}
The first analysis for bias we perform is whether certain recommenders have stronger tendencies to recommend movies with a particular production country. The underlying hypothesis is that recommenders based on a knowledge graph derived from a Wikipedia in a particular language will have a tendency to also recommend more movies from a production country where that language is spoken (e.g., the recommender based on German DBpedia could have a stronger tendency to recommend German or Austrian movies).
\begin{table}[t]
    \centering
    \begin{tabular}{|l|r|r|}
         \hline
         \textbf{Country} & \textbf{\# of movies} & \textbf{\# of ratings} \\
         \hline
         USA & 1,679 & 622,946 \\
         UK & 267 & 92,470 \\
         France & 127 & 27,362 \\
         Germany & 69 & 21,170 \\
         Italy & 62 & 10,887 \\
         Canada & 46 & 12,367 \\
         Australia & 30 & 13,330 \\
         Japan & 26 & 5,718 \\
         Spain & 16 & 3,813 \\
         Mexico & 15 & 6,790 \\
         \hline
    \end{tabular}
    \caption{Top 10 production countries in the dataset}
    \label{tab:top10_countries}
\end{table}

Table~\ref{tab:top10_countries} shows the top 10 production countries in the dataset. It can be observed that the dataset is heavily skewed towards movies from the USA and, to the lesser extent, UK, whereas other production countries only play a minor role.
\begin{table}[t]
    \centering
    \begin{tabular}{|l||r|r|r|r|r||r|}
         \hline
         \textbf{Country/KG} & \textbf{de} & \textbf{fr} & \textbf{it} & \textbf{ru} & \textbf{en} & \textbf{$c_e$} \\
         \hline
         USA & 0.728 & 0.750 & 0.762 & 0.761 & \textbf{0.782} & 0.744 \\
         UK & \textbf{0.136} & 0.143 & 0.098 & 0.091 & 0.108 & 0.110 \\
         France & 0.028 & 0.030 & 0.036 & \textbf{0.037} & 0.026 & 0.033 \\
         Germany & 0.012 & 0.018 & 0.012 & 0.030 & \textbf{0.034} & 0.025 \\
         Italy & \textbf{0.016} & 0.009 & 0.013 & 0.009 & 0.009 & 0.013 \\
         Canada & 0.020 & 0.009 & \textbf{0.021} & 0.005 & 0.006 & 0.015 \\
         Australia & 0.017 & 0.010 & 0.013 & 0.008 & \textbf{0.020} & 0.016 \\ 
         Japan & 0.006 & 0.005 & \textbf{0.012} & 0.004 & 0.006 & 0.007 \\
         Spain & 0.006 & 0.004 & \textbf{0.006} & 0.002 & 0.005 & 0.005 \\
         Mexico & 0.004 & 0.001 & 0.005 & \textbf{0.006} & 0.002 & 0.008 \\
         \hline
    \end{tabular}
    \caption{Fraction of recommendations for different production countries by knowledge graph. $c_e$ denotes the expected fraction based on the prevalence in the dataset.}
    \label{tab:country_recommendations}
\end{table}

Table~\ref{tab:country_recommendations} shows the fraction of movies from the top 10 production countries recommended by the systems based on the different knowledge graphs. We can see that the massive skew of the dataset towards US movies is also reflected in the results: except for the US movies, the majority of recommendations is below the expected value $c_e$.

Furthermore, it can be observed that, although significant differences in the behavior exist, there is no clear pattern of the form that follows the above mentioned hypothesis. Except for movies from the US and Australia, the peak of recommendations is always observed for a KG which is not in the respective language. For example, the fraction of German movies recommended by the system based on the English DBpedia is almost three times as high as the one based on the German DBpedia. Also, for other languages, the patterns are different: the highest fraction of French movies is recommended by the system based on the Russian KG, the highest fraction of Italian movies is recommended by the system based on the German KG, and so on.

\subsection{Bias for Genres}
In a second analysis, we inspect another possible bias induced by the different KGs, i.e., the bias to recommend movies from particular genres. Table~\ref{tab:top10_genres} shows the top 10 genres in the dataset.

\begin{table}[t]
    \centering
    \begin{tabular}{|l|r|r|}
         \hline
         \textbf{Genre} & \textbf{\# of movies} & \textbf{\# of ratings} \\
         \hline
         Drama & 721 & 174,635 \\
         Comedy & 562 & 184,700 \\
         Action & 351 & 133,342 \\
         Thriller & 320 & 74,457 \\
         Romance & 244 & 44,784 \\
         Horror & 225 & 22,700 \\
         Science Fiction & 183 & 52,648 \\
         Adventure & 172 & 36,827 \\
         Children's & 130 & 10,316 \\
         Crime & 115 & 7,621 \\
         \hline
    \end{tabular}
    \caption{Top 10 genres in the dataset}
    \label{tab:top10_genres}
\end{table}

\begin{table}[t]
    \centering
    \begin{tabular}{|l||r|r|r|r|r||r|}
         \hline
         \textbf{Genre/KG} & \textbf{de} & \textbf{fr} & \textbf{it} & \textbf{ru} & \textbf{en} & \textbf{$c_e$} \\
         \hline
         Drama & \textbf{0.198} & 0.170 & 0.187 & 0.172 & 0.190 & 0.162 \\
         Comedy & 0.191 & 0.192 & \textbf{0.207} & 0.198 & 0.166 & 0.168 \\
         Action & 0.089 & 0.010 & 0.074 & \textbf{0.129} & 0.112 & 0.123 \\ 
         Thriller & 0.072 & 0.086 & \textbf{0.097} & 0.088 & 0.084 & 0.095 \\
         Romance & 0.073 & 0.055 & \textbf{0.081} & 0.080 & 0.052 & 0.071 \\
         Horror & 0.043 & 0.050 & 0.044 & 0.043 & \textbf{0.053} & 0.043 \\
         Science Fiction & 0.055 & 0.045 & 0.044 & \textbf{0.056} & 0.053 & 0.073 \\
         Adventure & 0.053 & 0.045 & 0.053 & \textbf{0.070} & 0.049 & 0.063 \\
         Children's & 0.041 & \textbf{0.053} & 0.052 & 0.026 & 0.046 & 0.031 \\
         Crime & 0.029 & 0.039 & 0.025 & 0.044 & \textbf{0.045} & 0.038 \\
         \hline
    \end{tabular}
    \caption{Fraction of recommendations for different genres by knowledge graph. $c_e$ denotes the expected fraction based on the prevalence in the dataset.}
    \label{tab:genre_recommendations}
\end{table}

Table~\ref{tab:genre_recommendations} shows the recommendations based on the different knowledge graphs for the top 10 genres. Here, we can again observe some interesting deviations. The recommender based on the Russian DBpedia has a tendency towards action, science fiction, and adventure movies, while the one based on the Italian DBpedia tends to recommend more movies from the comedy, thriller, and romance genres. 
Those findings partially correlate with studies on the popularity of particular genres in different countries. In \cite{moviePopularity}, the authors discuss that, e.g., action movies are more popular in Russia than in English-speaking or European countries, and that comedy movies are more popular in Italy. Hence, it is likely that a local Wikipedia community in those countries will put more emphasis on editing movies in the respective genre in Wikipedia, which then leads those movies being more and better represented in the corresponding language-specific DBpedia, and ultimately a stronger bias of the recommender system based on that knowledge graph towards that genre.

\subsection{Specific Performance Differences}
The observation that recommenders based on different knowledge graphs expose biases towards particular genres also leads us to looking at the problem from a different angle. In particular, we want to analyze if recommenders based on different knowledge graphs work better or worse for single genres. To that end, we created partitions of our dataset by movie genre, and ran the recommender systems on those partitions. Overall, runs for ten different genres were performed.

\begin{table}[t]
    \centering
    \begin{tabular}{|l||r|r|r|r|r|}
         \hline
         \textbf{Genre/KG} & \textbf{de} & \textbf{fr} & \textbf{it} & \textbf{ru} & \textbf{en} \\
         \hline
         Drama & 0.040 & \textbf{0.045} & 0.034 & 0.030 & 0.040 \\
         Comedy & \textbf{0.078} & 0.067 & 0.055 & 0.053 & 0.068 \\
         Action & 0.091 & \textbf{0.114} & 0.089 & 0.080 & 0.105 \\ 
         Thriller & 0.083 & \textbf{0.085} & 0.061 & 0.064 & 0.080 \\
         Romance & 0.038 & 0.046 & 0.036 & 0.043 & \textbf{0.056} \\
         Horror & 0.073 & 0.072 & 0.066 & 0.040 & \textbf{0.082} \\
         Science Fiction & 0.101 & \textbf{0.124} & 0.106 & 0.080 & 0.095 \\
         Adventure & 0.090 & \textbf{0.115} & 0.093 & 0.097 & 0.082 \\
         Children's & \textbf{0.209} & 0.146 & 0.176 & 0.064 & 0.200 \\
         Crime & 0.097 & 0.098 & 0.084 & \textbf{0.121} & 0.099 \\
         \hline
    \end{tabular}
    \caption{Performance (F1) of recommenders for different genres by knowledge graph.}
    \label{tab:genre_performance}
\end{table}

The results are depicted in table~\ref{tab:genre_performance}. We can observe that there are rather strong differences between the genre-specific recommender performance. The French DBpedia, which has also been identified as the best source of background knowledge above, yields superior results for half of the genres. On the other hand, also the Russian DBpedia, which shows the worst overall performance, outperforms all other recommender systems on the crime genre. 

The differences on the individual genres are sometimes marginal, but for some genres (e.g., horror, children's), the best performing system can achieve results which are twice or even thrice as high as those which perform worst. This shows that there is no one-size-fits-all solution, and that the exploration of different knowledge graphs for a particular task and domain is at least as beneficial as the exploration of algorithmic alternatives.

\section{Conclusion and Future Work}
\label{sec:conclusion}
In this paper, we have conducted a comparative study of recommender systems based on different knowledge graphs, particularly: versions of DBpedia, based on Wikipedia in different languages. The experiment design has been chosen in a way that a basic recommendation strategy was chosen and fixed, and five different underlying knowledge graphs were used. The results show that there are considerable differences in preferences of the recommenders. Particularly, we analyzed production countries and genres, but our method is generally applicable to other categorical variables as well (e.g., gender of producer or director, low, medium and high budget, etc.).

The second major observation is that despite overall trends, not all knowledge graphs are equally well suited for particular recommendation tasks. When building a recommender system for movies from a particular genre, the globally best performing knowledge graph might not be the one which performs best locally on a given task. Here, we argue that the choice of a knowledge graph -- which is usually fixed upfront in most related works -- should be treated equally, if not even more important as fine tuning algorithms.

The problem of fixing a knowledge graph upfront is not limited to recommender systems. Knowledge graphs have been suggested to be used in other fields as well, such as explainable AI \cite{lecue2020role}, data interpretation \cite{paulheim2012generating}, or social media analysis \cite{piao2016exploring}. Like for recommender systems, biases induced by the choice of a particular knowledge graph have not been researched to a large extent here.

In the future, we see a few interesting directions to pursue. One of those is the extension of the analysis both to other domains, such as music or book recommendations, as well as the inclusion of further categorical variables, such as biases towards male or female authors, or black or white musicians. 

The inclusion of further knowledge graphs in studies like these is also an interesting area. With the advent of more cross-domain knowledge graphs, such as Wikidata \cite{vrandevcic2014wikidata}, CaLiGraph \cite{heist2019uncovering}, and DBkWik \cite{hertling2018dbkwik} we assume that each of those comes with its very own coverage biases, and a setup like the one discussed in this paper would be a way of systemically investigating the possible impact of such biases on downstream applications. Furthermore, it is an open question whether combining information from different knowledge graphs is a suitable way of reducing the individual biases.

Finally, while we argue that the selection of a particular knowledge graph is at least as important as the selection and fine-tuning of a recommender algorithm, interaction effects between the two decisions must not be neglected. We assume that, while there is no one-size-fits-all solution either on the knowledge graph nor on the algorithm side, the sweet spot for an optimal solution might not just be the straight forward combination of the knowledge graph and algorithm which perform best in isolation.

\bibliography{references.bib}

\end{document}